\documentclass[aps,pra,twocolumn]{revtex4}
\usepackage{amsfonts}
\usepackage{color}
\usepackage{eepic}
\usepackage{epic}
\usepackage{ulem}
\usepackage{graphicx}
\usepackage{texdraw}
\begin{document}
\preprint{}
\title{Characteristics and benchmarks of entanglement of mixed states - the two qubit case}

\author{Shanthanu Bhardwaj$^{1,3}$}
\email{shanth@uchicago.edu}
\author{V. Ravishankar$^{1,2}$}
\email{vravi@iitk.ac.in;vravi@rri.res.in}
\affiliation{${}^1$Department of Physics, Indian Institute of Technology, Kanpur-208016, INDIA  }
\affiliation{${}^2$Raman research Institute, Sadashivanagar, Bangalore-560080, INDIA}
\affiliation{${}^3$Department of Physics, University of Chicago, Chicago, IL 60637, USA }
\date{\today}
\begin{abstract}
We propose that the entanglement of mixed  states is characterised properly in terms of a probability density function $\mathcal{P}_{\rho}(\mathcal{E})$.  There is a need for such a measure since the prevalent measures (such as \textit{concurrence} and \textit{negativity}) for  two qubit systems are rough benchmarks, and not monotones of each other.  Focussing on the two qubit states, we provide an explicit construction of $\mathcal{P}_{\rho}(\mathcal{E})$  and show that it is characterised by a  set of parameters, of which concurrence is but one particular combination.   $\mathcal{P}_{\rho}(\mathcal{E})$ is manifestly invariant under  $SU(2) \times SU(2)$ transformations.  It can, in fact,  reconstruct the state up to local operations - with the specification of at most four additional parameters.  Finally the new measure  resolves the controversy regarding the role of entanglement in quantum computation in NMR systems.
\end{abstract}
\maketitle

\section{Introduction}

Quantum entanglement is a unique resource for novel (nonclassical) applications such as quantum algorithms \cite{Shor:1994jg+Grover:1996rk}, quantum cryptography \cite{BBBSS+BB84}, and more recently, metrology \cite{nature}. Thus, it plays a pivotal role in quantum information theory.  It is also central to the study of the foundations of quantum mechanics \cite{Bell:1964kc+CHSH}. It is not surprising that  an abiding interest in  quantum entanglement persists to this date. 

Entanglement of a bipartite system in a pure state is unambiguous and well defined. In contrast, mixed state entanglement (MSE) is relatively poorly understood mainly because entanglement, as an observable (denoting a property of the state) cannot be represented by a linear operator in the Hilbert space. Although many criteria such as entanglement of formation and separability have been proposed, there is a realization \cite{BDSW:1996} that no single quantity can adequately represent the entanglement contained in  a mixed state. It may, therefore, be worthwile investigating whether a complete description of MSE is possible in a manner such that the current criteria emerge as particular, albeit useful benchmarks.

We propose in this paper a characterisation of MSE in terms of a suitably defined probability density for entanglement, $\mathcal{P}_{\rho}(\mathcal{E} )$. The proposal is operational for any bipartite system. In this work, we focus on  two qubit systems for which we fully implement the definition. We find that  $\mathcal{P}_{\rho}(\mathcal{E})$ is characterised by its points of non-analyticity of various orders which completely capture the information on MSE. This central result is employed to shed light on various aspects
and manifestations of MSE.

The plan of the paper is as follows. In the next section we review briefly the existing criteria of MSE, and their drawbacks. Section III proposes the new definition in terms of
a probability density function, which will be constructed fully for the two qubit case in section IV.
Section V discusses several examples illustrating the proposal, and we show how an existing
criterion - the concurrence- emerges as but one benchmark. In section VI we discuss
an interesting application, {\it viz.}, to the problem of entanglement in NMR quantum computation.  In section VII, we address  the  question of reconstructibility of the state given its entanglement density. We show that, unlike in the case of every other definition, our prescription allows for an almost complete reconstruction of the state, upto local $SU(2) \times SU(2)$ operations. Section VIII concludes with a summary and outlook.

\section{ Criteria and benchmarks of MSE}
In this section we review very briefly various criteria and definitions of MSE. It is not our purpose to provide an exhaustive description of all the definitions of MSE. We refer the reader to literature \cite{Bruss:JMP43} for details. Our intention is to merely provide a motivation and a proper setting for the new definition.

Consider a bipartite spin system in a pure state, $\vert \psi\rangle = \sum_{m_1,~m_2} c_{m_1m_2} \vert m_1m_2\rangle$, where the expansion is understood in any separable basis.
The entanglement in the state is unambiguously quantified by the  entropy carried by the  reduced density matrix of either of the subsystems, $S[\rho_r]$. 
In the particular case of two qubits, equivalent criteria (in the sense of  
being relative monotones) include
the degree of mixedness $1-Tr \rho_r^2$, the determinant $\vert \rho_r \vert$,  and the concurrence
${\cal C} = 2\vert c_{\uparrow\uparrow} c_{\downarrow\downarrow} - c_{\downarrow \uparrow}c_{\uparrow \downarrow}\vert$, 
 in writing which we  employ the equivalent
 and a convenient notation $\uparrow (\downarrow) \leftrightarrow \frac{1}{2} (-\frac{1}{2})$. Operationally speaking, it is necessary and sufficient
to measure a single quantity, the degree of polarisation $P_{1}=P_{2} \equiv P$ of either of the qubits. We note parenthetically 
 that a knowledge of $P$ allows a reconstruction of the parent state upto local $SU(2) \times SU(2)$ operations (LO). Equivalently, entanglement determines the state upto LO. Indeed, 
writing $\vert \psi \rangle = cos (\theta/2)  \vert \uparrow \uparrow\rangle + \sin (\theta/2) \vert \downarrow \downarrow\rangle$ in its canonical basis, we obtain the relation 
$P = | \sin \theta|$ which demonstrates the claim.

A description of MSE is not that straight forward, even in the two qubit case. It is not difficult to realize that there exists no single parameter that characterises MSE \cite{BDSW:1996}.  Nevertheless, a number of concepts and associated quantities have been introduced in an attempt to capture the entanglement content in a mixed state, the two prominent of them being separability \cite{Werner} and entanglement of formation (EOF) \cite{BDSW:1996, Wootters:1997id}. In addition, other concepts such as entanglement cost,  distillable entanglement, relative entropy of entanglement and entanglement witnesses have been introduced. It is instructive to look at the extent to which  the above mentioned definitions  (i) satisfy the requirements of entanglement, (iii) are quantifiable  and, (iii) are equivalent.

\begin{itemize}
	\item \textit{Compatibility with the requirements}: An entanglement measure is expected to satisfy  continuity, additivity, subadditivity, convexity,  and a nonincreasing nature under LO, together with classical communication (LOCC) \cite{Bruss:JMP43}. Much work has been done in checking for the compatibility of the above measures. It is found that (i)  distillable entanglement violates convexity \cite{Shor2} and that (ii) relative entropy of entanglement violates additivity \cite{Vollbrecht}. It has not been established  whether entanglement cost is compatible with continuity, and if  EOF is compatible with  additivity \cite{Bruss:JMP43}. In short, there seems to be no single measure which
        is consistent with all the above constraints.

	\item \textit{Quantifiability}: The measures listed above are quantifiable, at best, in a limited sense: Concurrence which quantifies EOF   is defined only for  a 2QS \cite{Hill:1997pf}, and its generalisation to higher spin systems is not available. Negativity as a measure of non-separability  is  a necessary and sufficient condition only for two qubit and qubit-qutrit systems \cite{Peres:1996dw, Horodecki:1997vt}; for higher spin systems,  it is only a necessary condition. Other operational criteria such as majorization \cite{Nielsen} and reduction \cite{Horodecki:PRA59} are, again, only necessary (but not sufficient) conditions for separability.  Entanglement cost and distillable entanglement have eluded any quantification so far.

	 \item \textit{Mutual equivalence}: All of the above criteria are equivalent only for a pure state.
          Concurrence and negativity are, for instance, not relative monotones and are hence inequivalent \cite{Grudka}: 
 States with the same concurrence can have differing negativities and vice-versa, although for any given state, its negativity is never greater than concurrence. 
\end{itemize}
To summarise, none  of  the above quantities can, by itself, capture fully the entanglement that is contained in a mixed state. This observation suggests strongly that a complete description of MSE requires more than a specification of a parameter. 

To further emphasise the need for a better description of MSE, we note that a pure state description is almost always an idealisation and that any future experimental realization of quantum information processes will be with
quantum systems in mixed states. An unsatisfactory understanding of MSE will reflect, in turn, an imprecise
appreciation of the non-classical features that render quantum information processing possible.
An explicit example is provided by NMR quantum computers \cite{Gershenfeld, Cory}. Here, the qubits are prepared experimentally
in what is known as a pseudo pure state
$$
\rho_{ps} = \frac{1}{4}(1-\epsilon)\mathcal{I} +\epsilon |\psi \rangle \langle \psi|
,
$$ 
where, $|\psi \rangle$ is a Bell state. The concurrence and the negativity of $\rho_{ps}$ survive if $\epsilon > \frac{1}{3}$, while experimentally, $\epsilon \approx 10^{-6}$ (see \cite{Gershenfeld}). 

 These states would be essentially classical if we employ concurrence as a criterion for MSE, and no quantum gate operation should be possible with these states since the sole feature that distinguishes a quantum system from its counterpart is entanglement \cite{Schrodinger+EPR}. Yet, notwithstanding the vanishing of this measure,  nontrivial nonclassical gate operations with up to eight qubits have been reported  \cite{Anil}. More recently, a 12-qubit pseudopure state has been reported for a weakly coupled NMR system \cite{12-qubit}. While one could entertain the possibility of QC without entanglement \cite{Braunstein:PRL83+Gurvits:PRA68}, it is perhaps more fruitful to unravel the sense in which the inherent entanglement -- not captured by EOF or nonseparability -- is a resource for QC in these systems. Thus, there is a clear need to go beyond the above mentioned benchmarks and attempt to obtain a more complete description.  We address this problem and propose an alternative definition of MSE in the next section. 

\section{Description of MSE by a  probability density function}
	
	\subsection{Motivation for the definition} The new definition of MSE which we propose differs from the existing criteria in that we describe MSE in terms of a probability density for the entanglement. To motivate the idea, we recall that a mixed state description is required when the system is an ensemble of quantum systems, each of  which is in a pure state \cite{comm2}.  Entanglement has a sharp value for each pure state, and it should be natural that MSE be described properly by a distribution defined over the  microstates. 

This task is, however, not 
 as straightforward as it might seem. Because of the principle of superposition, the ensemble description of a quantum system in a mixed state, as a weighted distribution over a set of pure states,  is {\em not} unique. Expressed equivalently, there is no way of knowing how a system has been `prepared', unless it is in a pure state for, only  a pure state $|\psi \rangle$ belongs to a unique  one dimensional projection  $\vert \psi\rangle \langle \psi \vert$, with an eigenvalue 1 \cite{Landau:QM3}.
	Thus, although MSE may be expected to acquire a statistical character, and be characterised by a suitably defined probability density function (PDF), care must be exercised  such that the PDF for a given  $\rho$ it is not an artefact of its resolution in terms of any particular incoherent superposition of pure states. As the first step in finding the way out, we consider the class of special systems whose density operators are projection operators. Note that both the pure states and the fully unpolarised state belong to this class.

	\subsection{Definition of PDF when $\rho$ is a projection} Consider the case $\rho=\frac{1}{M} \Pi_M$, where the projection operator $\Pi_M$ has  rank $M$. Let ${\mathcal H}(\Pi_M)$ be the subspace projected by $\Pi_M$. Observe that 
	for all $|\psi \rangle \in {\mathcal H}(\Pi_M)$, $\langle \psi|\rho|\psi \rangle =\rm{const}$, which merely expresses the fact that the probability density in the $M$ dimensional manifold  ${\mathcal H}(\Pi_M)$ is uniform. The density in the complementary subspace is, of course, zero. This statement is exact and does not depend on the eigenbasis (or any other set of states) chosen to expand $\rho$. The probability density for the entanglement associated with the subspace may be  defined thus:
	\newtheorem{definition}{Definition}
	\begin{definition}[PDF when $\rho$ is a projection] Let the state $\rho = \frac{1}{M} \Pi_M$ be a $M$ dimensional projection operator. The probability density function for entanglement of this state is given by
	
	\begin{eqnarray} \label{eq:PDF_defn}
	\mathcal{P}_ {\Pi_M}(\mathcal{E}) = \frac{\int d\mathcal{H}_{\Pi_M} \,
	\delta(\mathcal{E_{\psi}} - \mathcal{E})}{\int d\mathcal{H}_{\Pi_M}}.
	\end{eqnarray}
	where $d\mathcal{H}_{ \Pi_M}$ is the volume  measure for the manifold $\mathcal{H}_{\Pi_M}$.
	\end{definition}
	To fix the volume measure in (\ref{eq:PDF_defn}), we observe that the group of automorphisms  $G$ of the subspace ${\mathcal H}(\Pi)$  leaves $\rho$ invariant. The measure should naturally be invariant under this group action and is, therefore, intimately related to the Haar measure of $G$. Indeed, let ${\mathcal H}(\Pi)$ be generated by the group action on any reference state $|\psi_0 \rangle$. Any state  $|\psi \rangle \in \mathcal{ H}(\Pi)$ can be obtained by the action of some $ g \in G$: $|\psi \rangle = g |\psi_0 \rangle$. Let $H$ be the stabilizer group of the ray associated with the reference state.  The measure $ d\mathcal{H}_{\Pi}$ is simply obtained by the Haar measure for $G$, after factoring out the Haar measure for $H$ \cite{group}. Since the Haar measure is invariant under the group action, and  pure state entanglement is invariant under under LO, it follows that the PDF is  invariant under LO. 

The extension of the pure state 
 entanglement (one dimensional projections) to states which are higher dimensional projections has thus turned out  to be straightforward and unambiguous. As we shall see in the explicit case of 2QMS which we study in detail, they have  a rich structure which can nevertheless be captured by specifying a few parameters which are invariant under LO. It  remains to further extend the definition to mixed states which are not projections. We take that up in the next subsection. 
	\subsection{PDF for any mixed state} To extend the above definition to mixed states without any restriction, we adopt the  guiding principle that two states which are close to each other should possess ``similar'' entanglement densities. For example, the entanglement of a state with distinct but nearly equal eigenvalues, should not differ from the entanglement of a completely unpolarised system. To accomplish this, we write  $\rho$ as a weighted  sum of projection operators $\Pi_M$ which satisfy the following property. Let $\mathcal{H}(\Pi_M)$ be the subspace (of $\textrm{dim} M$ ) projected by $\Pi_M$. We then require that $\mathcal{H}(\Pi_M) \subset \mathcal{H}(\Pi_{M+1});~ M=1, \cdots N-1$, where $N$ is the dimension of $\rho$. In terms of these nested projections $\Pi_M$,  we define the following:
	\begin{definition}[PDF for a mixed state] 
Let a state $\rho$, be resolved in  terms of nested projection operators as $\rho = \sum_{M=1}^N \omega_M \Pi_M$, with $\Pi_M$ satisfying the normalisation $\sum_M\omega_M =1$. The probability density function (PDF) for the entanglement of  $\rho$  is given by 
	\begin{eqnarray} \label{eq:PDFG_defn}
	\mathcal{P}_{\rho}(\mathcal{E}) = \sum_{M=1}^N \omega_M \mathcal{P}_{\Pi_M}(\mathcal{E})
	\end{eqnarray}
	where the PDF for a projection is defined in (\ref{eq:PDF_defn}).
	\end{definition}

	The definition given above is unambiguous since the weights can be easily determined in terms of the eigenvalues 
 of $\rho$. Let $\lambda_i^{\downarrow}$ be the eigenvalues of $\rho$, arranged in a nonincreasing order, belonging to the respective  eigenstates  $|\psi_i \rangle$. The eigenstates are not unique if the eigenvalues are degenerate, but they are of no consequence to us here. We first write the trivial identity 
	\begin{eqnarray}\label{eq:resolution}
	\nonumber \rho &=& (\lambda_1 - \lambda_2)\Pi_1 +  (\lambda_2 - \lambda_3)\Pi_2 + \\
	\nonumber & &  \cdots (\lambda_{N-1} - \lambda_N)\Pi_{N-1} +  \lambda_N\Pi_N \\
	& &  \equiv \sum_{M=1}^N \Lambda_M \Pi_M,
	\end{eqnarray}
	where the projections $\Pi_M = \displaystyle\sum_{j=1}^{M} |\psi_j\rangle \langle \psi_j|,~M=1, \cdots N$, satisfy the nestedness condition stated above. The weights  $\omega_M$ in (\ref{eq:PDFG_defn}) are easily read off as $\omega_M =\Lambda_M/\lambda_1$.

 With this identification, we see that the nonnegative vectors  $\Omega$ and $\Lambda$, defined by $\Omega =(\omega_1, \cdots, \omega_N) \equiv \lambda_1\Lambda$  have  natural, but rather different interpretations.
       The norm of $\Lambda$ is a measure of the purity of the state,  and lies in the range $[1, 1/N]$, the limiting cases corresponding to the pure and the completely mixed states respectively. The norm of $\Omega$ represents, on the other hand, the degree of projection onto a subspace. Thus, 
$\parallel\Omega\parallel$ takes its maximum value, 1 when $\rho$ is a pure projection. In any case, the form of $\rho$ in (\ref{eq:resolution}) demonstrates the assertion made above, {\it viz.}, that if a set of eigenvalues are close to each other, the state is then predominantly in the subspace spanned by their respective eigenstates, with only  a small spill over to the individual states. In the other case when an eigenvalue is much larger than the other, the spill over to the projection to higher dimensional subspaces is small. These observations establish the physical viability of the definition.

	We remark that the definition of MSE is valid for any bipartite system, and is operational in the sense that it can, in principle, always be evaluated. The entanglement distribution is governed by the invariant Haar measure associated with 
 the group of automorphisms of each subspace, as also the entanglements of the pure states belonging to it. Since they are invariant under LO, their structure cannot be 
 arbitrary. Thus {\it e.g.}, the PDF for a  $N-1$ dimensional projection will be characterised by a single parameter \textendash $~$ the entanglement of the pure state orthogonal to the subspace. Postponing an investigation to higher spins to a future work, we now implement the above definition to the most important case in quantum information theory, {\it viz.}, the two qubit system.

\section{PDF for a two qubit spin system}  
Two qubit systems are the most important from the view point of applications, and also because of the extensive theoretical analyses that they have received. We  focus our attention exclusively on 2QS in the rest of the paper. We  (a) analyse entanglement in states which are pure projections, and (b) their extension  to general states, (c) illustrate the distribution in a number of examples, (d) discuss the role of concurrence, (e) the problem of reconstructing the state given the PDF for entanglement, and the (f) reconciliation of NMR QC with entanglement. As our pure state measure, we choose  concurrence defined in the introduction. As pointed out, this choice does not amount to any loss of generality since all the measures of pure state entanglement are monotones of each other.

We first consider the special class of states, $ \rho = \frac{1}{M}\Pi_M;~M=1, \cdots 4$. The spectrum consists of only two eigenvalues, zero and $1/M$ 
 with respective degeneracies $4-M$ and $M$. The two limiting cases $d=1,4$ correspond to the completely polarised (pure states), and the completely unpolarised (mixed states) respectively. Each of the above cases will be analysed in detail. First the simplest of them all, {\it viz}., a pure state. 
	
	\subsection{One dimensional projections - the pure states} The Haar measure for the case $\rho =|\phi\rangle\langle\phi|$ is trivial since the group of automorphisms is given by the subgroup consisting only of the identity element. Thus, the PDF has the form 
	$$
	\mathcal{P}_{1}(\mathcal{E}) = \delta(\mathcal{E} - \mathcal{E}_{\phi}),
	$$
	in terms of the entanglement of $\vert \phi\rangle$. The PDF has a support only at $\mathcal{E}_{\phi}$, and the entanglement is characterised by  a single number. Note that any other choice of pure state entanglement simply rescales ${\mathcal{E}_{\phi}} \rightarrow \mathcal{E}^{\prime}_{\phi}$, in  a monotonic manner. The form of the PDF is unaffected. It may also be noted that the PDF determines the one dimensional projection upto LO. 
	
	\subsection{ Two dimensional projection - $\rho = \frac{1}{2}\Pi_2$} This particular class of states has the richest and the most interesting entanglement distribution.  Since the definition of PDF in (\ref{eq:PDFG_defn}) takes care of the normalisation through the group volume factor, we pay no attention to the trace  factor $\frac{1}{M}$ hence forth. The form of the PDF crucially depends on the nature of the subspace $\mathcal{H}(\Pi_2)$. Suppose that $\mathcal{H}(\Pi_2)$ is spanned by the basis $\{\vert m_1 m_2 \rangle,~\vert m_1 m_2^{'} \rangle \}$. Without any further computation, we see that every state $\psi \in \mathcal{H}(\Pi_2)$ is separable, giving a PDF which vanishes everywhere, except at $\mathcal{E}=0$. It is not difficult to see that the above statement holds for all subspaces related to the specified subspace by local operations. {\it i.e.}, $SU(2) \times SU(2)$ transformations.  Such an equivalence under LO  is valid for other PDF as well. It is, there
 fore, necessary and sufficient to study PDF for  $\mathcal{H}(\Pi_2)$ which belong to inequivalent classes under LO.  To that end, we construct a canonical basis in $\mathcal{H}(\Pi_2)$ by freely employing LO. \\

	\noindent{\it Canonical basis in $\mathcal{H}(\Pi_2)$}: Let  $\vert \psi \rangle \in \mathcal{H}(\Pi_2)$. Let $\vert \chi_1\rangle,~ \vert\chi_2\rangle$ be orthonormal and span $\mathcal{H}(\Pi_2)$. We have, 
	\begin{equation}\label{eq:basis}
	\vert \psi \rangle = \cos{\frac{\theta}{2}} e^{i\phi / 2} \vert \chi_1 \rangle + \sin{\frac{\theta}{2}} e^{- i\phi / 2} \vert \chi_2 \rangle,
	\end{equation}
	where $0 \le \theta \le \pi,~ 0 \le \phi \le 2\pi$.  The Haar measure is simply read off as $d\mathcal{H} = \sin \theta d\theta d\phi$. \\
		
	\noindent We  assert that in any $\mathcal{H}(\Pi_2)$, there  is a state which is separable.

	\noindent {\it Proof}: The demonstration is straightforward. Let $\vert \chi_1 \rangle,~\vert \chi_2 \rangle$ be an  orthonormal basis in $\mathcal{H}(\Pi_2)$. Let the entanglement of $|\chi_1 \rangle$, $\mathcal{E}_{\chi_1}=|\sin \alpha|$. Its canonical form is then given by
	$$
	|\chi_1 \rangle = \cos \frac{\alpha}{2} |\uparrow \uparrow \rangle + \sin \frac{\alpha}{2} |\uparrow \downarrow \rangle,
	$$
	whence 
	\begin{eqnarray}
	|\chi_2 \rangle = a\big(- \sin \frac{\alpha}{2} |\uparrow \uparrow \rangle +\cos \frac{\alpha}{2}
	|\downarrow \downarrow \rangle \big) \\
	+b | \uparrow \downarrow \rangle +c|\downarrow \uparrow \rangle \nonumber
	\end{eqnarray}
	with the condition $|a|^2 +|b|^2+ |c|^2 =1$.  Let us expand $|\psi \rangle$ in the above basis employing (\ref{eq:basis}). The condition that $\mathcal{E}_{\psi}=0$ yields the quadratic equation in $z= \tan( \frac{\theta}{2}) exp(i\phi)$:
	$$
	(a^2\frac{ \sin \alpha}{2}- bc)z^2 -az + \frac{ \sin \alpha}{2} =0.
	$$
	whose solutions are always physical, by virtue of the bijective mapping between the points on a sphere and the complex plane. 

	Let the separable state be chosen as  a basis state and be brought to its canonical form $|\eta_1\rangle =(1, 0, 0, 0)\}$ in writing which we have ordered the basis states as $\{|\uparrow \uparrow \rangle,~|\uparrow \downarrow \rangle,~|\downarrow \uparrow \rangle,~|\downarrow \downarrow \rangle$. Employing the residual LO (which leave $\vert \eta_1 \rangle$ invariant), we may write the orthogonal basis vector as $ |\eta_2 \rangle =(0, x, y, z=\sqrt{1-x^2-y^2})$, where $x, y, z \ge 0$.  The subspace  $\mathcal{H}(\Pi_2)$ is, therefore, characterised by two non-negative parameters say, $x,y$. The PDF would also be characterised by the two parameters, and gets implicitly determined by (\ref{eq:PDF_defn}).
	
	\subsubsection{Determination of the PDF}

\begin{figure}[ht]
\setlength{\unitlength}{0.120450pt}
\begin{picture}(1800,1500)(0,0)
\footnotesize
\color{black}
\thicklines \path(370,249)(411,249)
\thicklines \path(1676,249)(1635,249)
\put(329,249){\makebox(0,0)[r]{$0$}}
\color{black}
\thicklines \path(370,483)(411,483)
\put(329,483){\makebox(0,0)[r]{$2$}}
\color{black}
\thicklines \path(370,717)(411,717)
\put(329,717){\makebox(0,0)[r]{$4$}}
\color{black}
\thicklines \path(370,950)(411,950)
\put(329,950){\makebox(0,0)[r]{$6$}}
\color{black}
\thicklines \path(370,1184)(411,1184)
\put(329,1184){\makebox(0,0)[r]{$8$}}
\color{black}
\thicklines \path(370,1418)(411,1418)
\put(329,1418){\makebox(0,0)[r]{$10$}}
\color{black}
\thicklines \path(370,249)(370,290)
\thicklines \path(370,1418)(370,1377)
\put(370,166){\makebox(0,0){$0$}}
\color{black}
\thicklines \path(631,249)(631,290)
\thicklines \path(631,1418)(631,1377)
\put(631,166){\makebox(0,0){$0.2$}}
\color{black}
\thicklines \path(892,249)(892,290)
\thicklines \path(892,1418)(892,1377)
\put(892,166){\makebox(0,0){$0.4$}}
\color{black}
\thicklines \path(1154,249)(1154,290)
\thicklines \path(1154,1418)(1154,1377)
\put(1154,166){\makebox(0,0){$0.6$}}
\color{black}
\thicklines \path(1415,249)(1415,290)
\thicklines \path(1415,1418)(1415,1377)
\put(1415,166){\makebox(0,0){$0.8$}}
\color{black}
\thicklines \path(1676,249)(1676,290)
\thicklines \path(1676,1418)(1676,1377)
\put(1676,166){\makebox(0,0){$1$}}
\color{black}
\color{black}
\thicklines \path(370,249)(1676,249)(1676,1418)(370,1418)(370,249)
\color{black}
\put(82,833){\makebox(0,0)[l]{\shortstack{$\mathcal{P(E)}$}}}
\color{black}
\put(1023,42){\makebox(0,0){$\mathcal{E}$}}
\color{magenta}
\thicklines
\dottedline[*]{30}(370,405)(1350,405)(1350,249)
\color{blue}
\thicklines
\dottedline[x]{50}(370,249)(410,256)(449,263)(489,270)(528,278)(568,285)(607,292)(647,300)(687,307)(726,315)(766,323)(805,332)(845,340)(884,349)(924,359)(964,368)(1003,379)(1016,382)

\dottedline[x]{50}(1016,382)(1056,393)(1096,405)(1135,418)(1175,432)(1214,447)(1254,464)(1293,483)(1333,504)(1373,529)(1412,559)(1452,595)(1491,641)(1531,703)(1570,795)(1610,956)(1636,1177)(1650,1394)
\color{red}
\thicklines \path(370,249)(370,249)(370,249)(370,249)(383,249)(396,249)(409,250)(422,251)(448,252)(461,255)(501,259)(514,262)(527,265)(540,266)(566,268)(579,269)(592,271)(605,274)(618,277)(631,279)(644,280)(657,281)(670,283)(683,286)(697,291)(710,296)(723,294)(736,300)(749,299)(762,302)(775,304)(788,306)(801,307)(814,310)(827,312)(840,315)(853,316)(866,319)(879,320)(892,323)(905,326)(919,328)(932,330)(945,334)(958,335)(971,339)(984,340)(997,345)(1010,347)(1023,350)(1036,354)
\thicklines \path(1036,354)(1049,357)(1062,360)(1075,364)(1088,367)(1101,371)(1114,375)(1127,379)(1141,384)(1154,388)(1167,393)(1180,398)(1193,403)(1206,410)(1219,416)(1232,422)(1245,429)(1258,437)(1271,446)(1284,455)(1297,465)(1310,477)(1323,491)(1336,506)(1350,524)(1363,547)(1376,575)(1389,615)(1402,682)(1415,853)(1428,726)(1441,659)(1454,624)(1467,602)(1480,585)(1493,572)(1506,561)(1519,552)(1532,544)(1532,249)(1545,249)(1558,249)(1572,249)(1585,249)(1598,249)(1611,249)(1624,249)(1637,249)(1650,249)(1663,249)

\color{black}
\thicklines \path(370,249)(1676,249)(1676,1418)(370,1418)(370,249)
\end{picture}

\caption{Some Typical probability density functions for $\Pi_2$.  Note the solid  curve, which shows all the features of $\mathcal{P}_{2}(\mathcal{E})$.  It has a cusp at $\mathcal{E}_{cusp} = 0.8$ and goes to zero at $\mathcal{E}_{max} = 0.89$. The step function is an extreme example, where $\mathcal{E}_{cusp} = 0$, and the other dotted curve, has $\mathcal{E}_{cusp} = \mathcal{E}_{max} = 1$}
\label{2-D}
\end{figure}
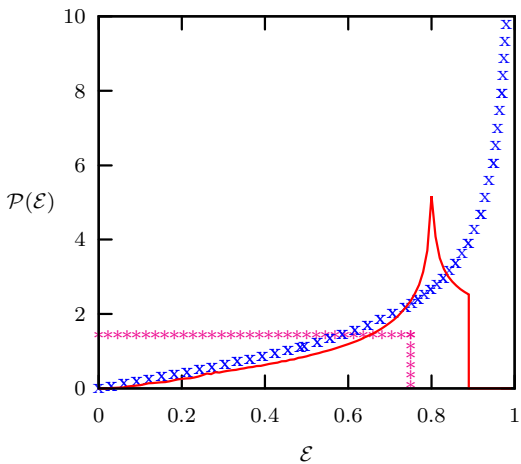
The determination of the distribution function is rather involved, and we give the details in the appendix.
Here, we present the results and discuss the salient features.

The generic form of the PDF for $\Pi_2$ is shown in FIG. \ref{2-D} (the solid curve).
We observe that it has three markers,  (i) $\mathcal{E}_{cusp}$, the entanglement at which the probability density diverges, as a cusp, (ii)$\mathcal{E}_{max}$, the maximum entanglement allowed, and (iii) $\mathcal{P}_2(\mathcal{E}_{max})$,  the probability density at $\mathcal{E}_{max}$.
In fact, any two of them suffice to characterise the PDF completely. One may  specify {\it e.g.},
 $(\mathcal{E}_{max},~ \mathcal{P}_2(\mathcal{E}_{max}))$, or equivalently, $(\mathcal{E}_{cusp},~ \mathcal{P}_2(\mathcal{E}_{max}))$ for characterising the curve. A straightforward computation establishes the relations

	\begin{eqnarray}\label{eq:Emax}
	\mathcal{E}_{max} &=& xy + \sqrt{z^2 + x^2y^2} \\ \label{eq:Ecusp}
	\mathcal{E}_{cusp} &=& \frac{z^2}{\mathcal{E}_{max}} = \mathcal{E}_{max} \cos \mu \\
	\nonumber \mu &=& \sin^{-1}\left(\frac{1}{\mathcal{E}_{max}\mathcal{P}_2(\mathcal{E}_{max})}\right) \\
	&=& \sin^{-1}\left( \frac{2 \sqrt{xy(xy\mathcal{E}_{max} + z^2)}}
	{\mathcal{E}_{max}^{3/2}} \right) \label{eq:mu}
\end{eqnarray}
which allow us to determine the parameters $x,y$ that define $\Pi_2$. $\mu$ is well defined by virtue of the inequality, $\mathcal{P}_2(\mathcal{E}_{max}) \geq  
1/\mathcal{E}_{max}$. 

The details of the nature of the PDF are presented in the Appendix, where the above
statements are proved. It is also shown that the PDF itself is an incomplete elliptic integral. For our purposes here, it is important that the curve is characterised by
the two locally invariant parameters $x,y$.

More importantly, we note that unlike with the other measures, the state itself can be  reconstructed up to LO.  For, we can reexpress $(x,y,z)$ in terms of the characteristics of the PDF thus: 
	\begin{eqnarray}
	&z = \sqrt{\mathcal{E}_{max} \mathcal{E}_{cusp}}  = \mathcal{E}_{max} \sqrt{\cos{\mu}}\\
	&x = \frac{1}{2}\left( \sqrt{(1 + \mathcal{E}_{max})(1 - \mathcal{E}_{cusp})} \right. \nonumber \\
	&\left. + \sqrt{(1 - \mathcal{E}_{max})(1 + \mathcal{E}_{cusp})} \right)\\ 
	&y= \frac{1}{2}\left( \sqrt{(1 + \mathcal{E}_{max})(1 - \mathcal{E}_{cusp})} \right.\nonumber \\
	&\left.- \sqrt{(1 - \mathcal{E}_{max})(1 + \mathcal{E}_{cusp})} \right) \\
	&xy = \frac{1}{2}\left( \mathcal{E}_{max} - \mathcal{E}_{cusp} \right)  = \mathcal{E}_{max} \sin^2(\mu/2)
	\end{eqnarray}

 The above equation expresses the result that the entanglement of a state which is a two dimensional
projection is completely characterised by its $SU(2) \times SU(2)$ invariant parameters, which are essentially two in number.
\subsubsection{Relation with concurrence and negativity}
We briefly discuss the status of two well known benchmarks, concurrence and negativity
in this description. It is, in fact, sufficient to consider concurrence since it bounds negativity from above. As a warm up, it is instructive to look at
two extreme cases which occur when $\mathcal{E}_{cusp} =0$ and $\mathcal{E}_{cusp} =\mathcal{E}_{max}$.
In the first case,
the PDF is a step function, terminating at some $\mathcal{E}_{max}$.
 In the second case, the density increases monotonically, diverging at $\mathcal{E}_{max}$
(see FIG. \ref{2-D}).
The relative abundance of the entangled states is more in the latter case.
One may \textit{per se}
expect that the associated concurrence should also be larger. Interestingly, however,   concurrence for a two
dimensional projection
is related to the new parameters by
\begin{eqnarray}
\mathcal{C}=(\mathcal{E}_{max} - \mathcal{E}_{cusp})/2.
\end{eqnarray}
Thus, contrary to  na\"{i}ve expectations, concurrence -- as a quantifier of entanglement of formation -- vanishes when 
$\mathcal{E}_{cusp}=\mathcal{E}_{max}$. In other words,  it is  sensitive not to the relative
abundance of the microstates at zero (or small entanglements) at all but,  to the
difference between $\mathcal{E}_{cusp}$ and $\mathcal{E}_{max}$.
 In any case,
$\mathcal{C}$ emerges as a particular benchmark of the probability density, describing it
only partially. 

We note that if $\rho = \Pi_3 ~~\rm{or}~~ \Pi_4$, then its concurrence vanishes identically.
By virtue of its convexity, we conclude that
  concurrence of any state $\mathcal{C}_{\rho}$, obeys the inequality
$$
\mathcal{C}_{\rho} \leq (\lambda_1 - \lambda_2) \mathcal{C}_{\Pi_1} +
(\lambda_2 - \lambda_3) \mathcal{C}_{\Pi_2}.
$$

	Incidentally,  entanglement distribution of a subspace $\mathcal{H}(\Pi_2^{c})$ orthogonal to $\mathcal{H}(\Pi_2)$ is the same as that of $\mathcal{H}(\Pi_2)$.  The proof of this statement is given in the Appendix.
	
	\subsection{\bf Three dimensional projection - $\rho = \frac{1}{3}\Pi_3$} We now move on to the case $\rho= \Pi_3$, whose PDF has a simpler structure. The
simplicity is afforded by the fact that $\Pi_3$ is completely characterised by its dual, $\vert \psi_{\perp} \rangle \perp \Pi_3$. Accordingly, its PDF is characterised by a single parameter $\mathcal{E}_{\perp}$, which is the entanglement of the orthogonal state $\vert \psi_{\perp} \rangle$. 

\begin{figure}[ht]
\definecolor{grey}{rgb}{.2, 0.2, .2}
\setlength{\unitlength}{0.120450pt}
\begin{picture}(1800,1500)(0,0)
\footnotesize
\color{black}
\thicklines \path(370,249)(411,249)
\thicklines \path(1676,249)(1635,249)
\put(329,249){\makebox(0,0)[r]{ 0}}
\color{black}
\thicklines \path(370,379)(411,379)
\thicklines \path(1676,379)(1635,379)
\put(329,379){\makebox(0,0)[r]{ 0.2}}
\color{black}
\thicklines \path(370,509)(411,509)
\thicklines \path(1676,509)(1635,509)
\put(329,509){\makebox(0,0)[r]{ 0.4}}
\color{black}
\thicklines \path(370,639)(411,639)
\thicklines \path(1676,639)(1635,639)
\put(329,639){\makebox(0,0)[r]{ 0.6}}
\color{black}
\thicklines \path(370,769)(411,769)
\thicklines \path(1676,769)(1635,769)
\put(329,769){\makebox(0,0)[r]{ 0.8}}
\color{black}
\thicklines \path(370,898)(411,898)
\thicklines \path(1676,898)(1635,898)
\put(329,898){\makebox(0,0)[r]{ 1}}
\color{black}
\thicklines \path(370,1028)(411,1028)
\thicklines \path(1676,1028)(1635,1028)
\put(329,1028){\makebox(0,0)[r]{ 1.2}}
\color{black}
\thicklines \path(370,1158)(411,1158)
\thicklines \path(1676,1158)(1635,1158)
\put(329,1158){\makebox(0,0)[r]{ 1.4}}
\color{black}
\thicklines \path(370,1288)(411,1288)
\thicklines \path(1676,1288)(1635,1288)
\put(329,1288){\makebox(0,0)[r]{ 1.6}}
\color{black}
\thicklines \path(370,1418)(411,1418)
\thicklines \path(1676,1418)(1635,1418)
\put(329,1418){\makebox(0,0)[r]{ 1.8}}
\color{black}
\thicklines \path(370,249)(370,290)
\thicklines \path(370,1418)(370,1377)
\put(370,166){\makebox(0,0){ 0}}
\color{black}
\thicklines \path(631,249)(631,290)
\thicklines \path(631,1418)(631,1377)
\put(631,166){\makebox(0,0){ 0.2}}
\color{black}
\thicklines \path(892,249)(892,290)
\thicklines \path(892,1418)(892,1377)
\put(892,166){\makebox(0,0){ 0.4}}
\color{black}
\thicklines \path(1154,249)(1154,290)
\thicklines \path(1154,1418)(1154,1377)
\put(1154,166){\makebox(0,0){ 0.6}}
\color{black}
\thicklines \path(1415,249)(1415,290)
\thicklines \path(1415,1418)(1415,1377)
\put(1415,166){\makebox(0,0){ 0.8}}
\color{black}
\thicklines \path(1676,249)(1676,290)
\thicklines \path(1676,1418)(1676,1377)
\put(1676,166){\makebox(0,0){ 1}}
\color{black}
\color{black}
\thicklines \path(370,249)(1676,249)(1676,1418)(370,1418)(370,249)
\color{black}
\put(82,833){\makebox(0,0)[l]{\shortstack{$\mathcal{P(E)}$}}}
\color{black}
\put(1023,42){\makebox(0,0){$\mathcal{E}$}}
\color{black}
\put(749,424){\makebox(0,0)[l]{$\mathcal{E}_{\perp} = 0.4$}}
\color{grey}
\put(892,1137){\vector(0,-1){693}}
\color{blue}
\thicklines \path(370,249)(370,249)(383,271)(396,294)(410,316)(423,339)(436,361)(449,384)(462,406)(476,428)(489,451)(502,473)(515,496)(528,518)(541,541)(555,563)(568,585)(581,608)(594,630)(607,653)(621,675)(634,698)(647,720)(660,742)(673,765)(687,787)(700,810)(713,832)(726,855)(739,877)(753,899)(766,922)(779,944)(792,967)(805,989)(819,1012)(832,1034)(845,1056)(858,1079)(871,1101)(884,1124)(898,1140)(911,1146)(924,1152)(937,1158)(950,1163)(964,1167)(977,1171)(990,1175)(1003,1178)(1016,1181)
\thicklines \path(1016,1181)(1030,1183)(1043,1185)(1056,1187)(1069,1188)(1082,1188)(1096,1188)(1109,1188)(1122,1187)(1135,1186)(1148,1184)(1162,1182)(1175,1179)(1188,1176)(1201,1172)(1214,1168)(1227,1163)(1241,1158)(1254,1153)(1267,1146)(1280,1140)(1293,1132)(1307,1124)(1320,1116)(1333,1107)(1346,1097)(1359,1087)(1373,1075)(1386,1064)(1399,1051)(1412,1038)(1425,1023)(1439,1008)(1452,992)(1465,975)(1478,957)(1491,938)(1505,918)(1518,896)(1531,872)(1544,847)(1557,820)(1570,791)(1584,760)(1597,725)(1610,686)(1623,642)(1636,592)(1650,530)(1663,449)(1676,249)
\color{black}
\thicklines \path(370,249)(1676,249)(1676,1418)(370,1418)(370,249)
\end{picture}
\caption{A Typical probability density for $\Pi_3$.  Note the point of discontinuity in the derivative at $\mathcal{E} = \mathcal{E}_{\perp}$}
\label{3-D}
\end{figure}
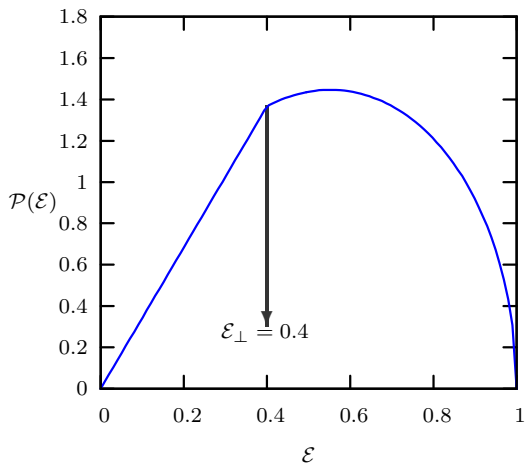

        Let 
 $\{\vert \chi_i \rangle \},~~i=1,2, 3$, be an orthonormal basis spanning the subspace $\mathcal{H} (\Pi_3)$, under consideration. 
	The integrating measure \cite{Byrd} may be conveniently written as $d\mathcal{H}_3= \sin 2\beta \sin 2\theta \sin^2 \theta d\alpha  d\beta  d\gamma  d\theta$, with the state  expanded   as $|\psi \rangle = \cos \theta |\chi_1 \rangle + e^{i (\alpha + \gamma)} \sin \theta \cos \beta |\chi_2 \rangle - e^{i (\alpha - \gamma)} \sin \theta \sin \beta |\chi_3 \rangle$. The ranges of  integration are given by $\theta, \beta \in [0, \frac{\pi}{2}]$ and $\alpha, \gamma \in [0, \pi]$.
        Using arguments similar to the ones employed for two dimensional subspaces,
	 one may, conveniently, choose two of the basis states, say $\chi_{1,2}$, to be separable;  by a suitable LO, they can be brought to the form $\vert \uparrow \uparrow \rangle,  | \downarrow\downarrow\rangle$. 

In this basis, the state 
$|\chi_3\rangle = c_1 |\uparrow \downarrow\rangle + c_2|\downarrow \uparrow\rangle$, has the same entanglement, $\mathcal{E}_{\perp}$, \newline as the state $|\psi_{\perp}\rangle = c_2^{*} |\uparrow \downarrow\rangle -c_1^{*}|\downarrow \uparrow\rangle$. In this canonical form the state looks like:
\[
	\rho = \left( \begin{array}{cccc}
\frac{1}{3} & 0 & 0 & 0 \\
0 & \frac{|c_1|^2}{3} & \frac{c_1c_2^{*}}{3} & 0 \\
0 & \frac{c_1^{*} c_2}{3} & \frac{|c_2|^2}{3} & 0 \\
0 & 0 & 0 & \frac{1}{3} \end{array} \right)
	\]
Here in terms of $c_1$, $c_2$, we have $\mathcal{E}_{\perp} = 2|c_1 c_2|$.

We have verified that the resulting probability density can be cast into  the simple form 
	\begin{eqnarray} 
	\mathcal{P}_3(\mathcal{E}) = \frac{2 \mathcal{E}}{\sqrt{1 - \mathcal{E}_{\perp}^2}} \cosh^{-1}{(\frac{1}{\mathcal{E}_{>}})}.
	\end{eqnarray} where $\mathcal{E}_{>} = \max (\mathcal{E}, \mathcal{E}_{\perp} )$.
	
	A typical curve for $\mathcal{P}_3(\mathcal{E})$ is shown in FIG. \ref{3-D}, which exhibits the required characteristic.  The curve possesses a discontinuity in its derivative at $\mathcal{E}_{\perp}$. Significantly, concurrence (being identically zero)  fails to distinguish different three dimensional projections, \textit{e.g.}, $\mathcal{E}_{\perp} =0 ~~\rm{or}~~ 1$, although their PDFs are vastly different. It simply picks up $\mathcal{E}=0$, at which
        the probability density, in fact, vanishes. 

        Since $\mathcal{P}_3(\mathcal{E})$ is characterised entirely by $\mathcal{E}_{\perp}$, it is clear that the state itself can be reconstructed upto LO.
	
	\subsection{\bf Full Hilbert Space} 
\begin{figure}[ht]
\setlength{\unitlength}{0.120450pt}
\begin{picture}(1800,1500)(0,0)
\footnotesize
\color{black}
\thicklines \path(411,249)(452,249)
\thicklines \path(1676,249)(1635,249)
\put(370,249){\makebox(0,0)[r]{$0$}}
\color{black}
\thicklines \path(411,379)(452,379)
\thicklines \path(1676,379)(1635,379)
\put(370,379){\makebox(0,0)[r]{$0.2$}}
\color{black}
\thicklines \path(411,509)(452,509)
\thicklines \path(1676,509)(1635,509)
\put(370,509){\makebox(0,0)[r]{$0.4$}}
\color{black}
\thicklines \path(411,639)(452,639)
\thicklines \path(1676,639)(1635,639)
\put(370,639){\makebox(0,0)[r]{$0.6$}}
\color{black}
\thicklines \path(411,769)(452,769)
\thicklines \path(1676,769)(1635,769)
\put(370,769){\makebox(0,0)[r]{$0.8$}}
\color{black}
\thicklines \path(411,898)(452,898)
\thicklines \path(1676,898)(1635,898)
\put(370,898){\makebox(0,0)[r]{$1$}}
\color{black}
\thicklines \path(411,1028)(452,1028)
\thicklines \path(1676,1028)(1635,1028)
\put(370,1028){\makebox(0,0)[r]{$1.2$}}
\color{black}
\thicklines \path(411,1158)(452,1158)
\thicklines \path(1676,1158)(1635,1158)
\put(370,1158){\makebox(0,0)[r]{$1.4$}}
\color{black}
\thicklines \path(411,1288)(452,1288)
\thicklines \path(1676,1288)(1635,1288)
\put(370,1288){\makebox(0,0)[r]{$1.6$}}
\color{black}
\thicklines \path(411,1418)(452,1418)
\thicklines \path(1676,1418)(1635,1418)
\put(370,1418){\makebox(0,0)[r]{$1.8$}}
\color{black}
\thicklines \path(411,249)(411,290)
\thicklines \path(411,1418)(411,1377)
\put(411,166){\makebox(0,0){$0$}}
\color{black}
\thicklines \path(664,249)(664,290)
\thicklines \path(664,1418)(664,1377)
\put(664,166){\makebox(0,0){$0.2$}}
\color{black}
\thicklines \path(917,249)(917,290)
\thicklines \path(917,1418)(917,1377)
\put(917,166){\makebox(0,0){$0.4$}}
\color{black}
\thicklines \path(1170,249)(1170,290)
\thicklines \path(1170,1418)(1170,1377)
\put(1170,166){\makebox(0,0){$0.6$}}
\color{black}
\thicklines \path(1423,249)(1423,290)
\thicklines \path(1423,1418)(1423,1377)
\put(1423,166){\makebox(0,0){$0.8$}}
\color{black}
\thicklines \path(1676,249)(1676,290)
\thicklines \path(1676,1418)(1676,1377)
\put(1676,166){\makebox(0,0){$1$}}
\color{black}
\color{black}
\thicklines \path(411,249)(1676,249)(1676,1418)(411,1418)(411,249)
\color{black}
\put(82,833){\makebox(0,0)[l]{\shortstack{$\mathcal{P(E)}$}}}
\color{black}
\put(1043,42){\makebox(0,0){$\mathcal{E}$}}
\color{blue}
\thicklines \path(411,249)(411,249)(424,269)(437,289)(449,308)(462,328)(475,348)(488,367)(500,387)(513,406)(526,426)(539,445)(552,464)(564,484)(577,503)(590,522)(603,541)(615,560)(628,578)(641,597)(654,616)(667,634)(679,652)(692,670)(705,689)(718,706)(730,724)(743,742)(756,759)(769,777)(782,794)(794,811)(807,828)(820,845)(833,861)(845,878)(858,894)(871,909)(884,925)(897,940)(909,955)(922,970)(935,985)(948,999)(960,1012)(973,1026)(986,1039)(999,1052)(1012,1064)(1024,1076)(1037,1087)
\thicklines \path(1037,1087)(1050,1098)(1063,1109)(1075,1119)(1088,1129)(1101,1139)(1114,1148)(1127,1156)(1139,1165)(1152,1172)(1165,1180)(1178,1187)(1190,1193)(1203,1199)(1216,1204)(1229,1209)(1242,1214)(1254,1217)(1267,1220)(1280,1223)(1293,1224)(1305,1225)(1318,1225)(1331,1224)(1344,1222)(1357,1219)(1369,1215)(1382,1209)(1395,1203)(1408,1195)(1420,1186)(1433,1175)(1446,1164)(1459,1151)(1472,1136)(1484,1120)(1497,1103)(1510,1084)(1523,1064)(1535,1041)(1548,1017)(1561,989)(1574,959)(1587,923)(1599,883)(1612,835)(1625,777)(1638,706)(1650,618)(1663,509)(1676,371)
\color{black}
\thicklines \path(411,249)(1676,249)(1676,1418)(411,1418)(411,249)
\end{picture}
\caption{The probability density $\mathcal{P}_4(\mathcal{E})$ for the entire Hilbert space.}
\label{Full-space}
\end{figure}
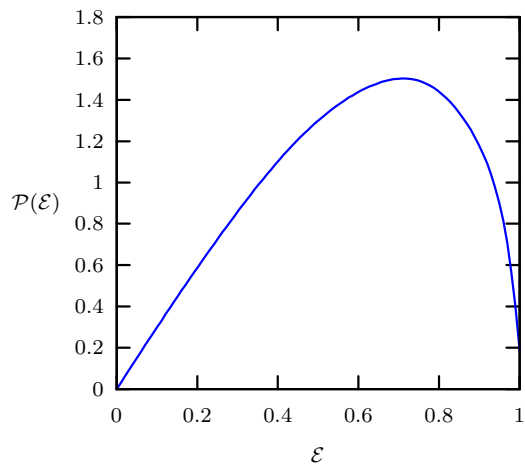

Lastly, we consider the full space $\mathcal{H}(\Pi_4)$, whose PDF is universal. This curve is obtained by using the Haar measure on $SU(4)$ \cite{Tilma:2002kf}. Note that the curve is smooth everywhere, as shown in FIG. \ref{Full-space}.

	\subsection{\bf A General Mixed State} The generalisation of the pure state entanglement
to higher dimensional projections has been accomplished so far. It remains to merely illustrate
the nature of PDF for entanglement when we have a superposition of nested projections, as given in
 (\ref{eq:resolution}). The entanglement density, as defined in 
 (\ref{eq:PDFG_defn}) does retain information on the contribution from the constituent nested projections, with appropriate weights $\omega_M$.  Indeed, each dimension produces a PDF with its
indelible characteristic. PDF for $\Pi_1$ is highly singular, being a Dirac delta distribution. PDF for $\Pi_2$ is less singular, but has a cusp as well as a step function discontinuity at $\mathcal{E}_{max}$. PDF
for $\Pi_3$ is smoother, posessing only a discontinuous derivative at $\mathcal{E}_{\perp}$. Finally, PDF for the fully unpolarised state, $\Pi_4$ is entirely smooth every where. Thus, the
definition of  mixed state entanglement given in  
(\ref{eq:PDFG_defn}) captures all the features
and stands vindicated. Recall that the weights have been so chosen that the continuity requirement
is maintained naturally.

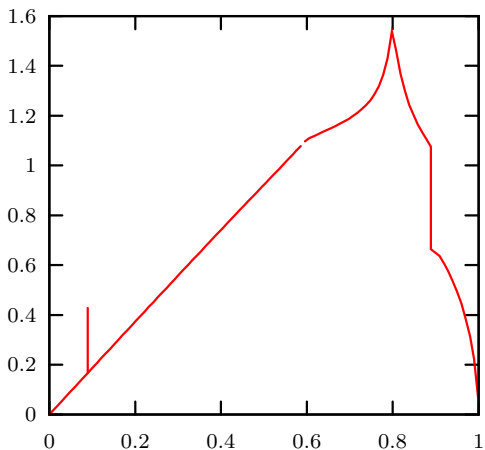
\begin{figure}[htb]
\setlength{\unitlength}{0.120450pt}
\begin{picture}(1800,1500)(0,0)
\footnotesize
\color{black}
\thicklines \path(328,166)(369,166)
\thicklines \path(1676,166)(1635,166)
\put(287,166){\makebox(0,0)[r]{$0$}}
\color{black}
\thicklines \path(328,323)(369,323)
\thicklines \path(1676,323)(1635,323)
\put(287,323){\makebox(0,0)[r]{$0.2$}}
\color{black}
\thicklines \path(328,479)(369,479)
\thicklines \path(1676,479)(1635,479)
\put(287,479){\makebox(0,0)[r]{$0.4$}}
\color{black}
\thicklines \path(328,636)(369,636)
\thicklines \path(1676,636)(1635,636)
\put(287,636){\makebox(0,0)[r]{$0.6$}}
\color{black}
\thicklines \path(328,792)(369,792)
\thicklines \path(1676,792)(1635,792)
\put(287,792){\makebox(0,0)[r]{$0.8$}}
\color{black}
\thicklines \path(328,949)(369,949)
\thicklines \path(1676,949)(1635,949)
\put(287,949){\makebox(0,0)[r]{$1$}}
\color{black}
\thicklines \path(328,1105)(369,1105)
\thicklines \path(1676,1105)(1635,1105)
\put(287,1105){\makebox(0,0)[r]{$1.2$}}
\color{black}
\thicklines \path(328,1262)(369,1262)
\thicklines \path(1676,1262)(1635,1262)
\put(287,1262){\makebox(0,0)[r]{$1.4$}}
\color{black}
\thicklines \path(328,1418)(369,1418)
\thicklines \path(1676,1418)(1635,1418)
\put(287,1418){\makebox(0,0)[r]{$1.6$}}
\color{black}
\thicklines \path(328,166)(328,207)
\thicklines \path(328,1418)(328,1377)
\put(328,83){\makebox(0,0){$0$}}
\color{black}
\thicklines \path(598,166)(598,207)
\thicklines \path(598,1418)(598,1377)
\put(598,83){\makebox(0,0){$0.2$}}
\color{black}
\thicklines \path(867,166)(867,207)
\thicklines \path(867,1418)(867,1377)
\put(867,83){\makebox(0,0){$0.4$}}
\color{black}
\thicklines \path(1137,166)(1137,207)
\thicklines \path(1137,1418)(1137,1377)
\put(1137,83){\makebox(0,0){$0.6$}}
\color{black}
\thicklines \path(1406,166)(1406,207)
\thicklines \path(1406,1418)(1406,1377)
\put(1406,83){\makebox(0,0){$0.8$}}
\color{black}
\thicklines \path(1676,166)(1676,207)
\thicklines \path(1676,1418)(1676,1377)
\put(1676,83){\makebox(0,0){$1$}}
\color{black}
\color{black}
\thicklines \path(328,166)(1676,166)(1676,1418)(328,1418)(328,166)
\color{black}
\color{red}
\thicklines
\path(449,297)(449,501)
\color{black}
\color{red}
\thicklines
\path(328,166)(328,166)(342,181)(355,195)(369,210)(382,225)(396,240)(410,254)(423,269)(437,284)(451,299)(464,313)(478,328)(491,343)(505,358)(519,372)(532,387)(546,402)(559,417)(573,431)(587,446)(600,461)(614,475)(628,490)(641,505)(655,519)(668,534)(682,549)(696,563)(709,578)(723,592)(736,607)(750,622)(764,636)(777,651)(791,665)(805,680)(818,694)(832,709)(845,723)(859,738)(873,752)(886,767)(900,781)(913,795)(927,810)(941,824)(954,838)(968,852)(982,867)(995,881)(995,881)(1009,895)(1022,909)(1036,924)(1050,938)(1063,952)(1077,967)(1091,981)(1104,996)(1118,1010)
\thicklines \path(1131,1025)(1145,1035)(1159,1041)(1172,1047)(1186,1054)(1199,1060)(1213,1066)(1227,1073)(1240,1080)(1254,1087)(1268,1095)(1281,1104)(1295,1114)(1308,1125)(1322,1138)(1336,1153)(1349,1173)(1363,1198)(1376,1233)(1390,1285)(1404,1370)(1417,1310)(1431,1237)(1445,1182)(1458,1140)(1472,1107)(1485,1078)(1499,1053)(1513,1030)(1526,1008)(1526,686)(1553,665)(1567,642)(1581,616)(1594,587)(1608,554)(1622,515)(1635,469)(1649,413)(1662,341)(1676,203)
\color{black}
\thicklines \path(328,166)(1676,166)(1676,1418)(328,1418)(328,166)
\end{picture}
\caption{The overall probability density $\mathcal{P}_4(\mathcal{E})$ for a typical mixed state, $\rho$, with eigenvalues \{0.385, 0.288, 0.231, 0.096\}. The features of the individual subspaces are vividly preserved. Note: The delta function is represented as a vertical line of height equal to its weight at the abscissa of its support}
\label{general-state}
\end{figure}
Before going on to discuss more interesting examples, we pause to illustrate explicitly how $\mathcal{P}_3(\mathcal{E})$ may be determined,  for
a rather arbitrarily chosen state. 
	Consider the density matrix $\rho$ with eigenvalues $\{ \lambda_i \} = \{ 0.385, 0.288, 0.231, 0.096 \} $ and the respective eigenvectors given by:
	\begin{eqnarray*}
	|\psi_1\rangle =& ( 0.998, 0.000, 0.031, 0.050)\,, \\
	|\psi_2\rangle =& ( 0.059, - 0.009, -0.528, -0.847)\,, \\
	|\psi_3\rangle =& ( 0.000, 0.924, 0.325, -0.202)\,, \\
	|\psi_4\rangle =& ( 0.000, 0.383, -0.784, 0.489)\,.
	\end{eqnarray*}
	
	Now, for the 1-D subspace, we simply have $\mathcal{E}_1 \approx 0.1$, and the weight associated with this delta function PDF is $(\lambda_1 - \lambda_4)/\lambda_1 \approx 0.25$.  Similarly, for the 2-D subspace spanned by $|\psi_1 \rangle, \vert \psi_2 \rangle$
        it is easy to verify that the canonical basis in $\Pi_2$ can be chosen to be
 $|\chi_1 \rangle = (1,0,0,0)$,  $|\chi_2 \rangle = (0, x, y, z) = (0, 0.00945, 0.5290, 0.8485)$.  From this it follows that $\mathcal{E}_{max} \approx 0.9$ and $\mathcal{E}_{cusp} \approx 0.8$ using (\ref{eq:Emax}), (\ref{eq:Ecusp}) and (\ref{eq:mu}).  Also, we can easily see that $\mathcal{E}_{\perp} \approx 0.6$ directly from $|\psi_4\rangle$.
	
	FIG. \ref{general-state} illustrates the PDF for this general case. Note how all the essential features stand out in the graph.
	
\section{\bf Examples} We proceed to study $\mathcal{P}(\mathcal{E})$ for states which are often
encountered, and also those which have a natural geometric structure. The exercise serves to
highlight the richness of the definition.

\subsection{Strongly separable states}As the first example we consider states which are strongly separable,  $\rho = \rho_1 \times \rho_2$. Conventional definitions of attribute no entanglement to this class of states.   We can very readily look at what form the PDF will take by making simple local rotations so that $\rho_1$ and $\rho_2$ are both written as 
$\frac{1}{2}(\mathbf{1} + k_{i}\sigma_z^{(i))})$ (where $i=1,2$ respectively), with $k_1, k_2 \ge 0$.  Thus $\rho$ acquires the form 
	
	\begin{eqnarray*}
	\frac{1}{4}\textrm{diag}(1 + k_1 + k_2 + k_1 k_2, 1 - k_1 + k_2 - k_1 k_2, \\
	1 + k_1 - k_2 - k_1 k_2, 1 - k_1 - k_2 + k_1 k_2)
	\end{eqnarray*}
	The eigenvector corresponding to the largest eigenvalue, in this case, is clearly $|\uparrow \uparrow \rangle$, therefore the 1-dimensional part of the PDF has no entanglement. The 2-D subspace is also separable since the eigenvector with the next largest eigenvalue is either $|\downarrow \uparrow \rangle$ or $|\uparrow \downarrow \rangle$, both of which form a separable subspace with $|\uparrow \uparrow\rangle$.  In other words, the PDF in $\Pi_2$ vanishes everywhere, except at $\mathcal{E}=0$. Therefore the first nonzero contribution to the PDF comes from the 3-dimensional subspace.  In this case too, it is the least possible contribution that is possible from a  3-D subspace because, $\mathcal{E}_{\perp} = 0$.  Thus, our measure of entanglement, the PDF, gives the minimum possible PDF to separable states. Thus the PDF
        is a simple superposition of FIG. \ref{3-D} and FIG. \ref{Full-space}. The entanglement vanishes of course if the state is pure.
 
\subsection{Purely vector polarised states} In the previous example, we had a nonvanishing tensor polarisation which was not independent of its vector polarisation. We now consider states which are purely vector polarised. These states are not factorisable. Further, they are never in a pure state.	
Writing $\rho = \frac{1}{4}(\mathbf{1} + \vec{p_1}\cdot\vec{\sigma_1} +  \vec{p_2}\cdot\vec{\sigma_2})$, we can bring it to the canonical form $\rho = \frac{1}{4}(\mathbf{1} + p_1\sigma^z_{1} +  p_2\sigma^z_{2})$.
An easy adaptation of the previous case shows again that the nonvanishing contribution comes from $\Pi_3$.
	
\subsection{Purely tensor polarised states} 
These states  come in three classes each of which we study below:
	\subsubsection{Pseudopure states} An important, but an easily analysable state is a pseudo pure state which is an incoherent superposition of a one dimensional projection and the
        projection operator for the full space. These states are employed in NMR QC, and unravelling their entanglement is not without interest. Pseudopure states have the form
 $\rho = \frac{1}{4}(\mathbf{1} + k \vec{\sigma_1} \cdot \vec{\sigma_2})$.  
Unlike in the previous cases, the sign of $k$ cannot be altered by a local transformation and it lies in the range $-1 \le k \le 1/3$. The eigenvalue decomposition of $\rho$ is given by
	\begin{eqnarray}\label{eq:pseudopure}
	\rho = \frac{1 + k}{4}\{ \Pi_{\uparrow \uparrow} +
          \Pi_{\downarrow \downarrow} + \Pi_B \}
 + \frac{1 - 3k}{4} \Pi_B^{\prime} \nonumber \\
\equiv \frac{1 -\epsilon}{4}\mathbf{1} +\epsilon \Pi_B^{\prime},,
	\end{eqnarray}
 where $\epsilon=-k$, and $\Pi_{\uparrow \uparrow}$, $\Pi_{\downarrow \downarrow}$ are the projection operators for the states $\vert \uparrow \uparrow\rangle$, $\vert \downarrow \downarrow \rangle$;  $\Pi_B$ and $\Pi_B^{\prime}$ are the projection operators for the respective Bell states
\begin{eqnarray*} 
\vert \psi_B \rangle=\frac{1}{\sqrt{2}}\{\vert \uparrow \downarrow \rangle + \vert \downarrow \uparrow\rangle\} \\ \vert\psi_B^{\prime} \rangle=\frac{1}{\sqrt{2}}\{\vert \uparrow \downarrow \rangle - \vert \downarrow \uparrow\rangle\}.
\end{eqnarray*}
The state is pure at the extremal value $k=-1$, and is the completely entangled singlet state.
It is completely unpolarised at $k=0$; at the other extremal value $k=1/3$,
 it is a three dimensional projection, orthogonal to the singlet state. Thus, for  $k<0$,
$\mathcal{E}_{\rho}$ gets a contribution from the Bell state ( the Dirac Delta has its support at $\mathcal{E}=1$) with a weight $\omega_1= \frac{-4k}{1-3k}$ and the full space with a weight $\omega_4=\frac{1+k}{1-3k}$. Similarly, when
 $k>0$,
its entanglement gets a contribution from the full space with a weight $\frac{1-3k}{1+k}$
and the three dimensional subspace (orthogonal to the singlet) with a weight $ \frac{4k}{1+k}$.
The curve corresponding to $\Pi_3$ is a straight line since $\mathcal{E}_{\perp}=1$.
There is no contribution from the two dimensional projection in either case. We take up a discussion of the import of this example to NMR QC in the next section.

\subsubsection{States of the form $\rho = \frac{1}{4}(\mathbf{1} + \vec{p}\cdot (\vec{\sigma_1}
 \times \vec{\sigma_2})$)} 
We can easily utilise local rotations to align $\vec{p}$ along the z-axis, thus converting the density matrix to the form:
	\[
	\rho = \frac{1}{4}\mathbf{1} + p(\sigma^{x}_{1} \otimes \sigma^{y}_{2} - \sigma^{y}_{1} \otimes \sigma^{x}_{2}) 
	\]
	\[
	\rho = \left( \begin{array}{cccc}
\frac{1}{4}  & 0 & 0 & 0 \\
0 & \frac{1}{4} & 2ip& 0 \\
0 & -2ip & \frac{1}{4} & 0 \\
0 & 0 & 0 & \frac{1}{4} \end{array} \right)
	\]

	Thus, we have the eigenvalues $\{\frac{1}{4} + 2p, \frac{1}{4}, \frac{1}{4}, \frac{1}{4}-2p\}$, with the respective eigenvectors: $\{\frac{|\uparrow \downarrow\rangle - i |\downarrow \uparrow\rangle}{\sqrt{2}}, |\uparrow \uparrow \rangle,|\downarrow \downarrow \rangle , \frac{|\uparrow \downarrow\rangle + i|\downarrow \uparrow\rangle}{\sqrt{2}}\}$. From the above structure, the PDFs for various subspaces and the associated weights may be easily obtained.
 For the one dimensional projection, we have the PDF and its associated weight given by
$$P_1(\mathcal{E})=\delta (\mathcal{E}-1); ~~\omega_1 = \frac{2p}{(\frac{1}{4} + 2p)}.
$$  
There is no contribution from the 2-D subspace, since $\omega_2=\lambda_2-\lambda_3 =0$. Considering the  3-D subspace, since $\mathcal{E}_{\perp}=1$, the probability density and the weights are read off as
$$          
P_{3}(\mathcal{E}) = 2 \mathcal{E};~~\omega_3 = \frac{2p}{(\frac{1}{4} + 2p)}.
$$ 
Interestingly, $\omega_1=\omega_3$.

  Thus the states belonging to the above class simply have PDFs that are essentially linear with a slope varying from $0$ to $1$, with a weighted $\delta$-function at $\mathcal{E} = 1$.

	\subsubsection{States with traceless symmetric tensor polarisation}
        Finally, we consider tensor polarised states in their most familiar -- the quadrupolar -- form
	 $\rho = \frac{1}{4}\{\mathbf{1} + A_{ij} \cdot ({\sigma^i_1}\otimes {\sigma^j_2})\}$, where $A_{ij}$ is a traceless symmetric matrix.  This matrix is diagonalisable by a local    $SU(2)\times SU(2)$ transformation.  We  bring the matrix to the form, $A = \textrm{diag}(A_{xx},A_{yy},-A_{xx}-A_{yy})$, where $A_{xx} \geq A_{yy} \geq 0$.  In this basis,  $\rho$ acquires the form,
	\[
	\rho = \left( \begin{array}{cccc}
\frac{1}{4} - \lambda & 0 & 0 & \mu \\
0 & \frac{1}{4} + \lambda & \lambda & 0 \\
0 & \lambda & \frac{1}{4} + \lambda & 0 \\
\mu & 0 & 0 & \frac{1}{4} - \lambda \end{array} \right)
	\]
	where $\lambda = A_{xx} + A_{yy}$, and $\mu = A_{xx} - A_{yy}$.  This gives us the eigenvalues 
	\[(\lambda_1, \lambda_2, \lambda_3, \lambda_4) = (\frac{1}{4} + 2\lambda, \frac{1}{4}, \frac{1}{4} - \lambda + \mu, \frac{1}{4} -\lambda - \mu),\]
	where $\lambda_1 \geq \lambda_2 \geq \lambda_3 \geq \lambda_4 $; and the corresponding eigenvectors are the Bell states: 
	\begin{eqnarray*}
	|\psi_1\rangle = \frac{1}{\sqrt{2}}(|\uparrow \downarrow\rangle + |\downarrow \uparrow \rangle), \\
	|\psi_2\rangle = \frac{1}{\sqrt{2}}(|\uparrow \downarrow\rangle - |\downarrow \uparrow \rangle), \\
	|\psi_3\rangle = \frac{1}{\sqrt{2}}(|\uparrow \uparrow \rangle + |\downarrow \downarrow\rangle), \\
	|\psi_4\rangle = \frac{1}{\sqrt{2}}(|\uparrow \uparrow \rangle - |\downarrow \downarrow\rangle).
	\end{eqnarray*}
        The rest of the analysis is straight forward. The probability density function and the associated weight for 
	 the Bell state $|\psi_1\rangle$ are simply read off as
          $$
	 \mathcal{P}_1(\mathcal{E}) = \delta(\mathcal{E}-1);~~\omega_1 = 
	 \frac{2\lambda}{\frac{1}{4} + 2\lambda};
          $$ 
          The  PDF for the two dimensional projection
         spanned by $|\psi_{1,2} \rangle$ also has a simple expression and the weight given by,  
         $$
         \mathcal{P}_2(\mathcal{E}) = 
	 \frac{\mathcal{E}}{\sqrt{1 - \mathcal{E}^2}};~~\omega_2 = \frac{\lambda - \mu}{\frac{1}{4} 
	 + 2\lambda}.
         $$ 
            Since $\vert \psi_4 \rangle$ is a Bell state, we obtain for the three dimensional subspace, 
         $$
         \mathcal{P}_3(\mathcal{E}) = 2\mathcal{E};~~\omega_3 = 
	 \frac{2\mu}{\frac{1}{4} + 2\lambda}.
          $$ 
           Finally the weight for the full space is given by $\frac{\frac{1}{4}-\lambda-\mu}{\frac{1}{4}+2\lambda}$.  It is remarkable that for this class of states, the details of the state manifest only in the weights. The density function for each dimension is itself universal. 
          Furthermore, we have constraints on the coefficients $\lambda+\mu \leq \frac{1}{4}$, and $\lambda \geq \mu \geq 0$.
	
          FIG. 5 illustrates the PDF for a typical state of this class.

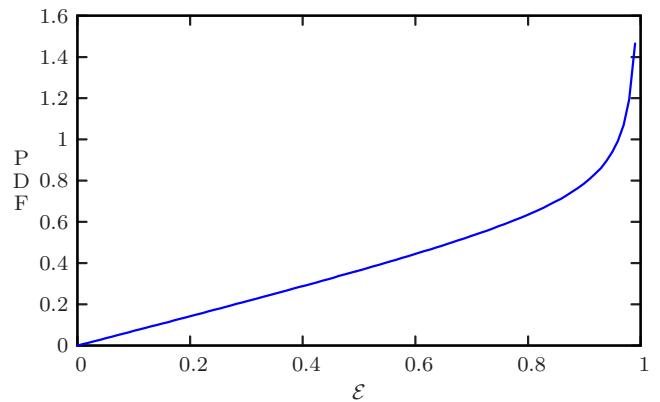
\begin{figure}[ht]
\setlength{\unitlength}{0.085pt}
\begin{picture}(3000,1800)(0,0)
\footnotesize
\thicklines \path(370,249)(411,249)
\thicklines \path(2876,249)(2835,249)
\put(329,249){\makebox(0,0)[r]{ 0}}
\thicklines \path(370,433)(411,433)
\thicklines \path(2876,433)(2835,433)
\put(329,433){\makebox(0,0)[r]{ 0.2}}
\thicklines \path(370,616)(411,616)
\thicklines \path(2876,616)(2835,616)
\put(329,616){\makebox(0,0)[r]{ 0.4}}
\thicklines \path(370,800)(411,800)
\thicklines \path(2876,800)(2835,800)
\put(329,800){\makebox(0,0)[r]{ 0.6}}
\thicklines \path(370,984)(411,984)
\thicklines \path(2876,984)(2835,984)
\put(329,984){\makebox(0,0)[r]{ 0.8}}
\thicklines \path(370,1167)(411,1167)
\thicklines \path(2876,1167)(2835,1167)
\put(329,1167){\makebox(0,0)[r]{ 1}}
\thicklines \path(370,1351)(411,1351)
\thicklines \path(2876,1351)(2835,1351)
\put(329,1351){\makebox(0,0)[r]{ 1.2}}
\thicklines \path(370,1534)(411,1534)
\thicklines \path(2876,1534)(2835,1534)
\put(329,1534){\makebox(0,0)[r]{ 1.4}}
\thicklines \path(370,1718)(411,1718)
\thicklines \path(2876,1718)(2835,1718)
\put(329,1718){\makebox(0,0)[r]{ 1.6}}
\thicklines \path(370,249)(370,290)
\thicklines \path(370,1718)(370,1677)
\put(370,166){\makebox(0,0){ 0}}
\thicklines \path(871,249)(871,290)
\thicklines \path(871,1718)(871,1677)
\put(871,166){\makebox(0,0){ 0.2}}
\thicklines \path(1372,249)(1372,290)
\thicklines \path(1372,1718)(1372,1677)
\put(1372,166){\makebox(0,0){ 0.4}}
\thicklines \path(1874,249)(1874,290)
\thicklines \path(1874,1718)(1874,1677)
\put(1874,166){\makebox(0,0){ 0.6}}
\thicklines \path(2375,249)(2375,290)
\thicklines \path(2375,1718)(2375,1677)
\put(2375,166){\makebox(0,0){ 0.8}}
\thicklines \path(2876,249)(2876,290)
\thicklines \path(2876,1718)(2876,1677)
\put(2876,166){\makebox(0,0){ 1}}
\thicklines \path(370,1718)(370,249)(2876,249)(2876,1718)(370,1718)
\put(82,983){\makebox(0,0)[l]{\shortstack{P\\D\\F}}}
\put(1623,42){\makebox(0,0){$\mathcal{E}$}}
\color{black}
\thicklines \path(370,1718)(370,249)(2876,249)(2876,1718)(370,1718)
\color{blue}
\thicklines
\path(370,249)(370,249)(395,256)(421,262)(446,269)(471,275)(497,282)(522,288)(547,295)(573,302)(598,308)(623,315)(648,321)(674,328)(699,335)(724,341)(750,348)(775,354)(800,361)(826,368)(851,374)(876,381)(902,388)(927,394)(952,401)(978,408)(1003,414)(1028,421)(1053,428)(1079,435)(1104,441)(1129,448)(1155,455)(1180,462)(1205,468)(1231,475)(1256,482)(1281,489)(1307,496)(1332,503)(1357,510)(1383,516)(1408,523)(1433,530)(1458,537)(1484,544)(1509,551)(1534,559)(1560,566)(1585,573)(1610,580)
\thicklines 
\path(1610,580)(1636,587)(1661,594)(1686,602)(1712,609)(1737,616)(1762,624)(1788,631)(1813,639)(1838,646)(1863,654)(1889,662)(1914,670)(1939,677)(1965,685)(1990,693)(2015,701)(2041,710)(2066,718)(2091,726)(2117,735)(2142,744)(2167,752)(2193,761)(2218,770)(2243,780)(2268,789)(2294,799)(2319,809)(2344,819)(2370,830)(2395,841)(2420,852)(2446,864)(2471,877)(2496,890)(2522,903)(2547,918)(2572,934)(2598,951)(2623,969)(2648,990)(2673,1013)(2699,1039)(2724,1071)(2749,1110)(2775,1160)(2800,1231)(2825,1345)(2851,1594)
\end{picture}
\caption{A typical quadrupolar state, with the values $\lambda = 0.15, \mu = 0.08$. Note that the weighted delta function at $\mathcal{E} = 1$, is not shown in the plot since the PDF goes to $\infty$ at this point.}
\end{figure}

        Before we conclude this section, we wish to add the cautionary remark that although
        the above analysis reveals entanglement in a host of states which are otherwise considered
        to be classical, it does not imply that all of
         them may be harnessed with equal facility. For instance, the entanglement of the uniform distribution cannot be accessed by standard gate operations which are unitary, and the corresponding PDF (FIG. \ref{Full-space}) has to be treated as a background. While more study is needed to
         discern the role of the entanglement density in various quantum information processes, there is one application of topical interest which we take up below.

\section{Entanglement in NMR Quantum Computation}
As an application of the new description, we address the issue of the role played by entanglement
in QC with NMR. 
NMR QC employs the so called pseudopure states which have the form 
$\frac{1 -\epsilon}{4}\mathbf{1} +\epsilon \Pi_B^{\prime}$
(\ref{eq:pseudopure}). It has been experimentally demonstrated that 
  the quantum logic  operations used in QC are implementable with NMR, and we know that no quantum logic operation is possible with classical states. Interestingly, concurrence and negativity vanish when $\epsilon <\frac{1}{3}$, while in experiments, $\epsilon \sim 10^{-6}$.

  This has led to a debate on the role of entanglement in NMR QC \cite{Braunstein:PRL83+Gurvits:PRA68} although, as we saw, experiments clearly show that such states cannot be completely classical.
 
  The PDF constructed for the pseudopure states in the previous section resolves this problem naturally. First of all its $\mathcal{P}_{\rho}(\mathcal{E})$ is given by a weighted Dirac Delta
  which is nonvanishing for all $\epsilon >0$, superposed on the background contribution from the
  uniform distribution. We know that
 the NMR signal is sensitive only to the pure component, the so called deviation deviation density matrix. Thus, although the uniform background is invariant under unitary operations,  the one dimensional fluctuation is not, allowing for non-trivial gate operations. In other words, NMR QC exploits the excess of entangled states over the unpolarised background as a resource, and this feature is correctly captured by the PDF of the state. Incidentally, this analysis also raises the interesting possibility of QC with more general pseudo projection states.

\section{Reconstructibility of $\rho$ from the PDF}
Lastly, we return to the issue of the reconstructibility of the state (up to LO). We have seen that when $\rho$ is a projection, the reconstructibility is  assured, by construction. 
When $\rho$ is more general, the reconstruction is somewhat partial;  we are not permitted to perform independent LO on various subspaces if the reconstruction is desired. Indeed, the action of $SU(2) \times SU(2)$ on $\rho$ produces an orbit of dimension six, characterised by nine invariants. The set of parameters which characterize the entanglement are seven in number, which may be chosen, for example, to be : $\{\mu_1, \mu_2, \mu_3, \mathcal{E}_{1}, \mathcal{E}_{cusp}, \mathcal{E}_{max}, \mathcal{E}_{\perp} \}$. Thus we need  
two additional parameters which would determine $\rho$. To understand their role, we note that geometrically, $\mathcal{P}_{\rho}(\mathcal{E})$ is invariant under independent Local operations $L_i$, acting on the subspaces $\Pi_i$, where $\Pi_i \subset \Pi_{i+1}$. If $\rho$ is to be unique up to a global LO, one needs the additional constraint $L_i = U L_{i}^{(0)}$, where $L_{i}^{(0)}$ may be chosen freely. Let us choose $L_2^{(0)} = \mathbf{1}$ (where $\mathbf{1}$ is the identity operator).  The nestedness condition, \textit{viz.}, that $|\psi_1\rangle \in \Pi_2$ and $|\psi_4\rangle \in \Pi_2^c$, entails that $L_1^{(0)}$ and  $L_3^{(0)}$ get specified by two parameters each. In fact we need only one parameter each to fix the states up to discrete ambiguities.  This is because from the $\mathcal{P}_{i}(\mathcal{E})$ we already  know the entanglement of the states, which fixes one parameter.  However, there remains a discrete ambiguity if only one of $\theta$ or $\phi$ is specified.

We make the above argument more explicit.
If we have $\Pi_2$ in its canonical form, it is spanned by $|\chi_1 \rangle$ and $|\chi_2\rangle$ given respectively as: $(1,0,0,0)$ and $(0,x,y,z)$.  Therefore, we can specify $|\psi_1\rangle = |\chi_1 \rangle \cos{\frac{\theta}{2}} e^{i\phi / 2} + |\chi_2 \rangle \sin{\frac{\theta}{2}} e^{- i\phi / 2}$ by giving the values of $(\theta, \phi)$. Similarly, $|\psi_{\perp} \rangle$ can be specified by $(\theta_{\perp}, \phi_{\perp})$ when it is expanded in the canonical basis of $\Pi_2^{c} = (\mathbf{1} - \Pi_2)$, given by $|\chi_1^{c} \rangle = (0,0,c/ \sqrt{c^2+b^2},-b/ \sqrt{c^2+b^2})$ and $|\chi_2^{c} \rangle = (0, \sqrt{c^2+b^2}, ab/ \sqrt{c^2+b^2}, ac/ \sqrt{c^2+b^2})$. The above construction completes the argument. It is noteworthy that  the
question of reconstructibility cannot even be raised with other criteria.\\

\section{Conclusion}In conclusion, we have shown that mixed state entanglement has a rich structure
and is properly described via a suitably defined probability density. We have explicitly implemented
the definition to the most important case, the two qubit systems, and shown how criteria such as concurrence emerge as specific bench marks. Their precise role in describing the entanglement is also  clarified. The role of entanglement in NMR QC is resolved and, for the first time, the issue of reconstructibility of the state discussed. Nevertheless, the study is incomplete since possible applications to  teleportation, quantum algorithms and error correcting codes are still to be explored.
The generalisation to higher spin systems would also provide a deeper and a better appreciation of
quantum information processes.

\section{Appendix}
The properties of the entanglement density $\mathcal{P}_{\rho}(\mathcal{E})$ for a two dimensional
projection $\Pi_2$ will be worked out in detail here.

Consider $\Pi_2$ first.
 Recall that we are considering the subspace spanned by $|\alpha\rangle = (1,0,0,0)$ and $|\beta\rangle = (0,x,y,z)$.   A general state $|\psi\rangle = \cos\frac{\theta}{2}e^{i\phi/2}|\alpha\rangle + \sin\frac{\theta}{2}|\beta\rangle $ has its entanglement given by
\begin{equation} \label{eq:ent}
\mathcal{E}^2 = |2(z \sin(\theta/2) \cos(\theta/2)e^{i\phi/2} + xy \sin^2(\theta/2) ) |^2. \\
\end{equation}
It follows from the above expression that the maximum entanglement allowed is given by
\begin{equation} \label{emax}
\mathcal{E}_{max} = xy +\sqrt{z^2+x^2y^2}
\end{equation}
It is further convenient to introduce the variable $\mu$ defined by
 $z = \mathcal{E}_{max} \sqrt{\cos{\mu}}$, and $xy = \mathcal{E}_{max} \sin^2(\mu/2)$. It follows that 
 the entanglemet of \textit{every} state in the subspace scales linearly in $\mathcal{E}_{max}$.  Therefore, we can write 
$$\left. \mathcal{P(E)} \right|_{(\mathcal{E}_{max}\,,\,\mu)} = \frac{1}{\mathcal{E}_{max}}\left. \mathcal{P}^{'}(\mathcal{E}/\mathcal{E}_{max}) \right|_{(\mathcal{E}^{'}_{max} = 1\,,\,\mu)}
$$ if $\mathcal{E} < \mathcal{E}_{max}$ and $\left. \mathcal{P(E)} \right|_{(\mathcal{E}_{max}\,,\,\mu)} = 0$ if  $\mathcal{E} > \mathcal{E}_{max}$ .  We shall now utilise this scaling and concentrate on studying the distribution for $\mathcal{E}_{max} = 1$.  At $\mathcal{E}_{max}$, $xy = (1-z^2)/2$ and 
\begin{eqnarray*}
\mathcal{E}^2(\theta, \phi) = z^2 \sin^2\theta + \left( \frac{1-z^2}{2} \right)^2(1 - \cos\theta)^2 + \\
2z \left( \frac{1-z^2}{2}\right) \sin\theta(1- \cos\theta)\cos\phi
\end{eqnarray*}
and 
\begin{eqnarray*}
\mathcal{P(E)} = \frac{1}{4\pi} \int_0^{\pi} \sin\theta\,d\theta \int_0^{2\pi} d\phi \,\delta\left(\mathcal{E} - \mathcal{E}(\theta, \phi)\right) \\
= \frac{\mathcal{E}}{2\pi} \int_0^{\pi} \sin\theta\,d\theta \int_0^{2\pi} d\phi \,\delta\left(\mathcal{E}^2 - \mathcal{E}^2(\theta, \phi)\right) \\
\end{eqnarray*}

\begin{figure}
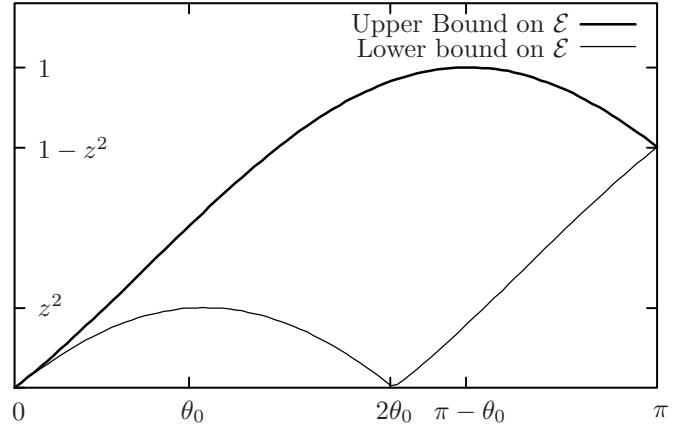

\begin{texdraw}
\normalsize
\ifx\pathDEFINED\relax\else\let\pathDEFINED\relax
 \def\QtGfr{\ifx (\TGre \let\YhetT\cpath\else\let\YhetT\relax\fi\YhetT}
 \def\path (#1 #2){\move (#1 #2)\futurelet\TGre\QtGfr}
 \def\cpath (#1 #2){\lvec (#1 #2)\futurelet\TGre\QtGfr}
\fi
\drawdim pt
\setunitscale 0.19
\linewd 3
\textref h:L v:C
\linewd 4
\path (154 90)(174 90)
\path (154 248)(174 248)
\path (1433 248)(1413 248)
\move (182 258)\htext{ $z^2$}
\path (154 568)(174 568)
\path (1433 568)(1413 568)
\move (182 568)\htext{ $1-z^2$ }
\path (154 728)(174 728)
\path (1433 728)(1413 728)
\move (182 728)\htext{ 1 }
\move (154 45)\textref h:C v:C \htext{ 0}
\move (501 45)\htext{ $\theta_0$}
\path (501 90)(501 110)
\path (501 856)(501 836)

\move (902 45)\htext{ $2\theta_0$}
\path (901 90)(901 110)
\path (901 856)(901 836)

\move (1052 45)\htext{ $\pi - \theta_0$}
\path (1052 90)(1052 110)
\path (1052 856)(1052 836)

\move (1433 45)\htext{ $\pi$}
\path (154 90)(1433 90)(1433 856)(154 856)(154 90)
\move (1259 814)\textref h:R v:C \htext{Upper Bound on $\mathcal{E}$}
\linewd 5
\path (1281 814)(1389 814)
\path (154 90)(167 100)(180 111)(193 121)(206 132)(219 143)
\cpath (232 155)(244 166)(257 178)(270 190)(283 202)
\cpath (296 214)(309 226)(322 238)(335 250)(348 263)
\cpath (361 275)(374 288)(387 301)(399 313)(412 326)
\cpath (425 339)(438 351)(451 364)(464 376)(477 389)
\cpath (490 401)(503 414)(516 426)(529 438)(542 451)
\cpath (554 463)(567 474)(580 486)(593 498)(606 509)
\cpath (619 520)(632 531)(645 542)(658 553)(671 563)
\cpath (684 573)(697 583)(710 593)(722 602)(735 611)
\cpath (748 620)(761 629)(774 637)(787 645)(800 652)
\cpath (813 660)(826 667)(839 673)(852 679)(865 685)
\cpath (877 691)(890 696)(903 701)(916 705)(929 709)
\cpath (942 713)(955 716)(968 719)(981 722)(994 724)
\cpath (1007 725)(1020 727)(1033 728)(1045 728)(1058 728)
\cpath (1071 728)(1084 727)(1097 726)(1110 725)(1123 723)
\cpath (1136 721)(1149 718)(1162 715)(1175 711)(1188 708)
\cpath (1200 703)(1213 699)(1226 694)(1239 688)(1252 683)
\cpath (1265 677)(1278 670)(1291 663)(1304 656)(1317 649)
\cpath (1330 641)(1343 633)(1355 625)(1368 616)(1381 607)
\cpath (1394 598)(1407 589)(1420 579)(1433 569)
\move (1259 769)\htext{Lower bound on $\mathcal{E}$}
\linewd 3
\path (1281 769)(1389 769)
\path (154 90)(154 90)(167 100)(180 110)(193 119)(206 128)
\cpath (219 137)(232 146)(244 154)(257 162)(270 170)
\cpath (283 178)(296 185)(309 191)(322 198)(335 204)
\cpath (348 210)(361 215)(374 220)(387 225)(399 229)
\cpath (412 233)(425 236)(438 239)(451 242)(464 244)
\cpath (477 246)(490 248)(503 249)(516 249)(529 250)
\cpath (542 249)(554 249)(567 248)(580 247)(593 245)
\cpath (606 243)(619 240)(632 237)(645 234)(658 231)
\cpath (671 226)(684 222)(697 217)(710 212)(722 207)
\cpath (735 201)(748 194)(761 188)(774 181)(787 174)
\cpath (800 166)(813 158)(826 150)(839 141)(852 132)
\cpath (865 123)(877 114)(890 104)(903 94)(916 96)
\cpath (929 106)(942 117)(955 127)(968 138)(981 150)
\cpath (994 161)(1007 173)(1020 184)(1033 196)(1045 208)
\cpath (1058 220)(1071 233)(1084 245)(1097 257)(1110 270)
\cpath (1123 282)(1136 295)(1149 308)(1162 320)(1175 333)
\cpath (1188 346)(1200 358)(1213 371)(1226 383)(1239 396)
\cpath (1252 408)(1265 421)(1278 433)(1291 445)(1304 457)
\cpath (1317 469)(1330 481)(1343 493)(1355 504)(1368 515)
\cpath (1381 526)(1394 537)(1407 548)(1420 559)(1433 569)
\path (154 90)(1433 90)(1433 856)(154 856)(154 90)
\end{texdraw}
\caption{The upper and lower bounds on $\mathcal{E}$ from inequalities (\ref{eq:upper_limit}) and (\ref{eq:lower limit}) are plotted above. } \label{Graph_of_bounds_on_E}

\end{figure}

Now we can do the $\phi$ integral pretty easily, and it leaves us with:
\begin{eqnarray*}
\mathcal{P(E)} = \frac{\mathcal{E}}{2\pi} \int_0^{\pi} \frac{\sin\theta\,d\theta}{2z \left( \frac{1-z^2}{2}\right) \sin\theta(1- \cos\theta)|\sin\phi_0|} \times 2 \\
\end{eqnarray*}
where $$\cos\phi_0 = \frac{\mathcal{E}^2 - z^2 \sin^2\theta - \left( \frac{1-z^2}{2} \right)^2(1 - \cos\theta)^2}{2z \left( \frac{1-z^2}{2}\right) \sin\theta(1- \cos\theta)}.$$  However, we need this solution for $\cos\phi_0$ to lie in $[-1,1]$.  Therefore:
\begin{eqnarray}
\nonumber \mathcal{E} &&< \left| z \sin\theta + \left( \frac{1-z^2}{2} \right)(1 - \cos\theta) \right| = U(\theta)\\
\Rightarrow \mathcal{E} &&< \left( \frac{1-z^2}{2} \right) - \left( \frac{1+z^2}{2} \right)\cos(\theta + \theta_0) \label{eq:upper_limit}
\end{eqnarray}
\begin{eqnarray} \nonumber \textrm{and }\mathcal{E} > \left| z \sin\theta - \left( \frac{1-z^2}{2} \right)(1 - \cos\theta) \right| = L(\theta) \\
\Rightarrow \mathcal{E} > 
	\left\{\begin{array}{cl}
		\left( \frac{1+z^2}{2} \right)\cos(\theta - \theta_0) - \left( \frac{1-z^2}{2} \right), & \theta < 2\theta_0 \\
		\left( \frac{1-z^2}{2} \right) - \left( \frac{1+z^2}{2} \right)\cos(\theta - \theta_0) , & \theta > 2\theta_0
	\end{array}\right. \label{eq:lower limit} 	
\end{eqnarray} where $\theta_0 = 2 \tan^{-1} z$. 

The integral for $\mathcal{P(E)}$ now becomes:
\begin{eqnarray*}
\frac{\mathcal{E}}{\pi} \int \frac{\sin\theta\,d\theta}{\sqrt{(\mathcal{E}^2 - L^2(\theta))( U^2(\theta)- \mathcal{E}^2)}} \\
\end{eqnarray*} where the integration is carried out over the region where inequalities (\ref{eq:upper_limit}) and (\ref{eq:lower limit}) are satisfied (see FIG. \ref{Graph_of_bounds_on_E}).  The denominator of the integrand goes to zero only at the boundaries, and both $U(\theta)$ and $L(\theta)$ have nonzero slopes almost everywhere (except $L(\theta)$ at $\theta_0$). Therefore, if we look near such a point, $\theta_b$, we see that :
\begin{eqnarray*}
L^2(\theta_b + \epsilon) - \mathcal{E}^2 &&= (\mathcal{E} + L'(\theta_b)\epsilon + \mathcal{O}(\epsilon^2))^2 - \mathcal{E}^2 \\
&&= 2\mathcal{E} L'(\theta_b) \epsilon + \mathcal{O}(\epsilon^2) 
\end{eqnarray*} thus near the points where the integrand blows up the behaviour is $\sim 1/\sqrt{\epsilon}$, which is convergent.  A special point to check at is when $\mathcal{E} = 1 - z^2$, then near $\theta = \pi$, both the terms in the denominator of the integral behave as $\sim 1/\sqrt{\epsilon}$, which is a $\sim 1/{\epsilon}$ behaviour, however, the $\sin\theta$ in the numerator also goes as $\sim {\epsilon}$, therefore the integral is convergent for this values as well.  Therefore, we are left to consider the case when $\mathcal{E} = z^2$.  In this case near $\theta_0$, the slope of $L(\theta)$ vanishes.  Therefore, for $\theta = \theta_0 + \epsilon$, $\mathcal{E}^2 - L^2(\theta) \sim \epsilon^2$.  Thus the integrand behaves as $\sim 1/\epsilon$, and we have a logarithmic divergence at $\mathcal{E} = z^2 = \cos\mu$.  This is the cusp in the PDF. This will also scale as $\mathcal{E}_{max}$ for values of $\mathcal{E}_{max} \neq 1$.  Thus $\mathcal{E}_{cusp} = \mathcal{E
 }_{max}\cos\mu$ as mentioned earlier.

It may be noted that the integral for the PDF can be recast into the form :
\begin{eqnarray}
\mathcal{P(E)} = \left\{\begin{array}{cl}
		 \int_{t_1}^{t_2} \frac{dt}{\sqrt{R(t)}}, \textrm{ if } \mathcal{E} > z^2\\
		\int_{t'_1}^{t'_2} \frac{dt}{\sqrt{R(t)}} + \int_{t'_3}^{t'_4} \frac{dt}{\sqrt{R(t)}} \textrm{ if } \mathcal{E} < z^2
	\end{array}\right. 
\end{eqnarray} where $R(t)$ is a polynomial of degree $4$ in $t$, if we make the substitution $t = \cos\theta$.  This is basically an incomplete elliptic integral (the limits $t_1,t_2 ...$ are obtained from the inequalities (\ref{eq:upper_limit}) \& (\ref{eq:lower limit})).

We lastly prove the result that the PDF for two complementary two dimensional projections are identical.
This follows from the fact that there is a bijective mapping from $\mathcal{H}(\Pi_2)$ to $\mathcal{H}(\Pi_2^{c})$ which preserves the Haar volume (in fact this map is an $SU(2) \times SU(2)$ transformation).  To demonstrate this we will once again choose our basis as $|\chi_1\rangle$ and $|\chi_2\rangle$ defined above.  Now, $\mathcal{H}(\Pi_2^{c})$ consists of all states orthogonal to $|\chi_1\rangle$ and $|\chi_2\rangle$.  We now construct a basis for $\mathcal{H}(\Pi_2^{c})$, as: $|\chi_1^{'}\rangle = (0, z/\sqrt{z^2+x^2},0,-x/\sqrt{z^2+x^2}) $ and  $|\chi_2^{'}\rangle = (0, xy/\sqrt{z^2+x^2},-\sqrt{z^2+x^2},zy/\sqrt{z^2+x^2})$.  It is easy to see that $|\chi_1^{'}\rangle$ is separable and $\mathcal{E}_{\chi_2^{'}} = \mathcal{E}_{\chi_2}$.  Furthermore, the entanglement of a general state, $|\psi\rangle = \alpha |\chi_1^{'}\rangle + \beta |\chi_2^{'}\rangle$, in the subspace is given by: $\mathcal{E}(\alpha,\beta) = 2| -\alpha\beta z - \beta^2 xy|$.  This is identical to!
  the entanglement of $\alpha |\chi_1\rangle + \beta |\chi_2\rangle$.  Since the entanglement of each state in the subspace is identical to that in $\mathcal{H}(\Pi_2)$ and the $SU(2)$ measure is the same, we will have the same PDFs in both these cases.  In fact the $SU(2) \times SU(2)$ transformation that takes $|\chi_1^{'}\rangle$ to $|\chi_1\rangle$, and $|\chi_2^{'}\rangle$ to $|\chi_2\rangle$ connects these two subspaces.


\begin{references}
\bibitem{comm2} The system behaves classically only when the measuring apparatus fails to see the quantum
        features in the microstates. Thus, a completely unpolarized spin half state is to be distinguished
        from an ensemble of unbiased coins. In other words, classicality emerges when coherent superposition in the microstates is not detectible.
\bibitem{group} 
      The relevant measure is, strictly speaking, the volume element of the coset space $G/H$,
where $H$ is the stabilzer group of the reference state, $\mathcal{H}(\Pi_M)$. Put more simply, the volume element corresponding to the stabilizer group gets dropped. This is illustrated for a two dimensional projection, where, the $SU(2)$ measure $sin \theta d\theta d\phi d\psi \rightarrow sin \theta d\theta d\phi$, since the stabilizer, $U(1)$ has the haar measure $d\psi$.
\end{references}
\end{document}